\documentclass[aps,prb,floats, eqsecnum,amsmath,graphicx,euscript,
amsmath,amssymb,twocolumn,showpacs,10pt]{revtex4}
\usepackage[dvips]{graphicx}
\begin{document}

\title{Cotunneling mechanism of single-electron shuttling}
\author {Guy Z. Cohen}
\author{V. Fleurov}
\author{K. Kikoin}
\affiliation{Raymond and Beverly Sackler Faculty of Exact Sciences,
School of Physics and Astronomy, Tel-Aviv University, Tel-Aviv 69978
Israel}

\begin{abstract}
The problem of electron transport by means of a dumbbell shaped
shuttle in strong Coulomb blockade regime is solved. The electrons
may be shuttled only in the cotunneling regime during the time spans
when both shoulders of the shuttle approach the metallic banks. The
conventional Anderson-like tunneling model is generalized for this
case and the tunneling conductance is calculated in the adiabatic
regime of slow motion of the shuttle. Non-adiabatic corrections are
briefly discussed.
\end{abstract}
\maketitle
\section{Introduction}

The idea of inducing charge transport in nanodevices by means of
special type of time-dependent manipulations with the parameters of
a device was formulated theoretically as adiabatic pumping
\cite{Thou} and nano-electromechanical shuttling \cite{Gorel}
mechanisms. Both phenomena are related to charge transfer through a
nanoobject under perturbation $V(x,t)$ periodic both in time $t$ and
a coordinate $x$. In case of pumping, $x$ is not necessarily a
spatial coordinate. It may be, e.g., the point on the Coulomb
diamond diagram for the tunnel conductance $G(v_{g_1}, v_{g_2})$ in
double quantum dots, which circumscribes closed trajectories as a
function of the gate voltages $v_{g1}$ and $v_{g2}$ applied to two
dots \cite{Poth,Swit,Zor}. Not only charge but also spin density may
be pumped within this paradigm.\cite{Watson} Electron-electron
interaction plays essential part in the adiabatic pumping. In
particular, the resonance Kondo-like tunneling is essential for the
spin-charge separation of the pumping component of tunnel current
(see, e.g., Ref. \onlinecite{Aono} for discussion of this problem
and references therein to other theoretical investigations of
adiabatic pumping in quantum dots).

In pumping devices the nanoobject is immovable, and single-electron
transport is induced by time-dependent control parameters. In the
nanoelectromechanical shuttling (NEMS) the charge and spin transport
occurs due to an interplay of electronic and mechanical degrees of
freedom: a nanoobject (or its parts) move periodically in real space
between the electrodes and transport electrons from one electrode to
another in the course of this motion. In the "classical"
NEMS\cite{Gorel,Gorev} the shuttling instability develops because
the charge accumulation in the shuttle induced  by electron
tunneling results in an electromechanical instability of the shuttle
at a sufficiently large bias voltage and periodic mechanical motion
of the shuttle arises in the system. Mechanical motion of the
shuttle may be also induced in the "dot" geometry, where a
nano-island is attached to a flexible pillar or
cantilever.\cite{Erbe,Kot2004} As a rule, shuttling in these devices
is substantially nonadiabatic.

Usage of flexible nanowires in the shuttling circuits opens new
exciting possibilities. The flexural vibration of a suspended
nanowire in combination with a scanning tunnel microscope working as
one of the two electrodes may be used for realization of a charge
transport in the shuttling regime.\cite{Leroy,Sheht09} On the other
hand the nano-island may be attached to such a string. Then
mechanical motion of a shuttle will be provided by excitation of
vibrational eigenmodes of this string.\cite{Koenig} If the Coulomb
blockade in the island is strong enough, the single electron
shuttling regime may be realized under certain experimental
conditions.\cite{Koendis} Although the shuttle motion induced by the
string flexure is slow in comparison with characteristic tunneling
rate, the adiabaticity is usually violated when the shuttle
approaches the electrodes and moves away.

In this  paper we consider the single electron transport in
suspended NEMS geometry. We study  the shuttle shaped like a
dumbbell (double quantum dot) rocking periodically around a fixed
axis. Such turnstile transports electrons by means of cotunneling
mechanism. First, we formulate  the problem of cyclic tunneling in
general terms. Second, we calculate the periodic time-dependent
conductance in the adiabatic regime. Finally, the role of
non-adiabaticity will be discussed qualitatively.

\section{Model}

As is known, cotunneling mechanism is a cornerstone of the Kondo
regime of tunnel conductance through quantum dots under strong
Coulomb blockade. The cotunneling amplitude $J_{lr}$, which arises
in the second order perturbation theory in tunneling integrals
$V_{l,r}$ between the dot and the left and right metallic leads is
exponentially weak, $J_{lr}\sim V_l V_r/\Delta$ ($\Delta$ is the
ionization energy of quantum dot). Due to formation of many-particle
Abrikosov-Suhl resonances at the Fermi level, this amplitude
approaches unity at low temperatures ($T \to 0$),\cite{GR,Ng} and
the zero bias anomaly (ZBA) in quantum conductance shows up in the
Coulomb diamond window.\cite{Kogla}

Shuttling provides another enhancement mechanism for single electron
tunneling, which exists even in the absence of Kondo
screening.\cite{KKSV} This mechanism is an analog of the
Debye-Waller effect in neutron scattering intensity: the average
distance between the dot and the leads effectively reduces due to
periodic motion of the shuttle between the source and drain
electrodes. The exponential enhancement of effective tunneling
transparency is controlled by the ratio $\langle x^2 \rangle/x_0^2$,
where the numerator is the average mean square displacement of the
shuttle position relative to its static distance $x_0$ from the
leads (the device is supposed to be symmetric). Recently we have
described one more enhancement mechanism unrelated to Kondo effect,
which may be realized in half-metals with the energy gap for
minority spin carriers.\cite{CFG2} This mechanism may be evinced as
a finite bias anomaly in conductance, which arises due to opening of
a resonance tunneling channel for the minority spin carriers.

\begin{figure}[ht]\begin{center}
  \includegraphics[width=6cm,angle=0]{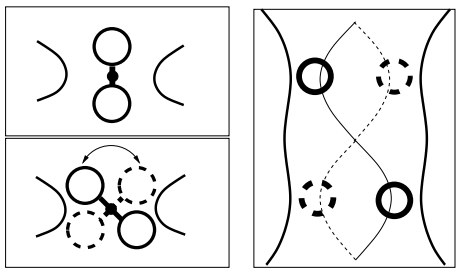}
  \caption{Left panel: Turnstile shuttle in a form of a dumbbell suspended on
  a twisting string oriented perpendicularly to the tunneling plane. Right plane:
  Turnstile shuttle in a form of double dot suspended on a string vibrating
  in the tunneling plane.}\label{f.0}
\end{center}
\end{figure}

In this paper we consider tunneling through a dumbbell shaped
shuttle. A schematic sketch of the device is presented in Fig.
\ref{f.0} (left panel).
We suppose that this shuttle is suspended on an elastic string
oriented perpendicularly to the figure plane, e.g., by means of
the method developed in Ref. \onlinecite{Koenig}. Another option
is to suspend two islands on a string where the standing wave
polarized in the tunneling plane is excited (Fig. \ref{f.0}, right
panel).

\begin{figure}[ht]\begin{center}
  \includegraphics[width=6cm,angle=0]{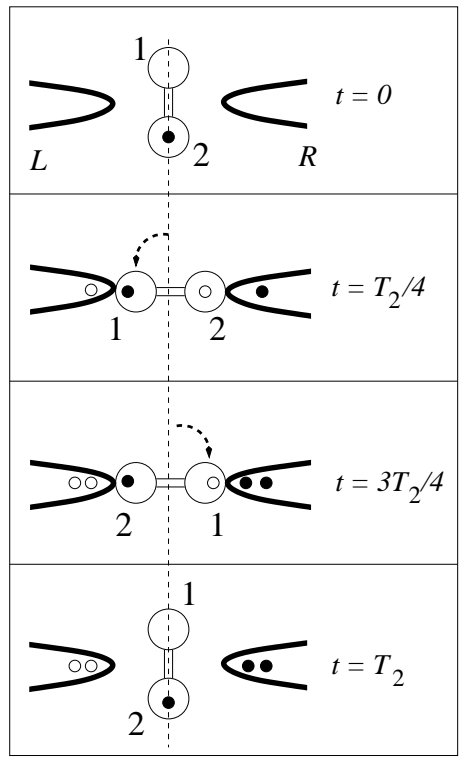}
  \caption{Cotunneling cycle of a turnstile shuttle}\label{f.2}
\end{center}
\end{figure}

Let us turn to the twisting turnstile as a more simple in practical
realization object. We consider the idealized case where the
equilibrium position with zero strain corresponds to the symmetric
orientation of the dumbbell relative to the edges. Then the torque
vibration mode may be excited in the string, which induces the
in-plane rotation of the dumbbell. When the rotation angle
approaches $\pi/2$, the distances between the dots and \textsl{both}
leads become small and the tunneling amplitude increases
simultaneously for the source and the drain lead (see Fig.
\ref{f.2}). If the Coulomb blockade is strong enough, the number of
electrons in the turnstile is fixed and an electron from the lead
can tunnel from the source into a turnstile only provided another
electron leaves the dot simultaneously. Thus effective elastic
cotunneling arises as a ZBA in the tunnel conductance with strongly
enhanced amplitude.

One may say that electrons tunnel between two banks in a shuttling
regime, where a carrier may be injected from a bank to the moving
island during the limited time span, when one of the two islands
is passing this bank. In our idealized geometry this condition is
satisfied by assuming that in the equilibrium position at rotation
angle $\varphi = 0$ tunneling is exponentially weak and electron
can jump onto the shuttle or leave it only at $\varphi \to \pm
\pi/2$. Second, the characteristic frequencies of string
vibrations are of the order of several MHz under the realistic
experimental conditions.\cite{Koenig} This means that the
mechanically driven shuttle moves slowly in comparison with the
characteristic times of electron motion, and the time dependent
processes are adiabatic for the most part of the shuttling period.
Adiabaticity may be violated only when the shuttle approaches the
metallic banks and tunneling amplitudes grow exponentially with
time reaching its maximum for $\varphi = \pm \pi/2$ and then again
decreasing to nearly zero value. Thus the torque vibrations of a
dumbbell-shape shuttle are transformed into periodically pulsing
tunnel amplitudes with completely synchronized source-shuttle and
shuttle-drain tunneling processes. Due to non-adiabatic character
of switching, this tunneling is accompanied by formation of
quasienergy levels, which contribute to the single-electron
shuttling.

To separate the single electron shuttling from Kondo
shuttling,\cite{KKSV} we consider here spinless carriers. This
approximation is valid provided the leads are fabricated from
half-metals, i.e. materials in which the Fermi surface is formed by
majority spin electrons, whereas the spectrum of minority spin
electrons is gapped \cite{CFG2}. The turnstile shuttle (TS) may be
described within a framework of time-dependent Anderson model with
the Hamiltonian
\begin{equation}\label{2.1}
H = H_{\rm sh} + H_{\rm L} + H_{\rm R}
+ H_{\rm tun}
\end{equation}
Here the first term is Hamiltonian of the moving TS
\begin{equation}\label{2.2}
H_{\rm sh} = \sum_{i=1,2} E_i n^{}_{i} + U_{12}n_1n_2;
\end{equation}
$n_i = d^\dag_i d^{}_i$ is the occupation operator for the electrons
in the wells of TS labeled as $i=1,2$; $E_i$ are the corresponding
discrete energy levels. Usually the time dependence is explicitly
included in the shuttle coordinate,\cite{Gorel,KKSV} because the
shuttle is driven by external electric field. We suppose that the
turnstile shuttle moves together with the acoustically excited
string, and the time variable appears only in the last term in the
Hamiltonian (\ref{2.1}) (see below). The second term in Eq.
(\ref{2.2}) is the electrostatic interaction between electrons
located in two potential wells. There is no need to introduce the
single site Coulomb blockade potential because the Pauli principle
forbids double occupation of any well. We assume that the coupling
$U_{12}$ is strong enough and the TS may be occupied only by one
electron located either in the level $E_1$ or in the level $E_2$.
The electrons in the left (\textrm{L}) and right (\textrm{R}) banks
are represented by the Hamiltonian
\begin{equation}\label{2.3}
H_{\rm L} + H_{\rm R} =\sum_{\beta=L,R} \sum_{\kappa}
\varepsilon_{\beta\kappa}a^\dag_{\beta\kappa}a^{}_{\beta\kappa}
\end{equation}
with quasicontinuous spectra $\varepsilon_{l\kappa}$. The last term
in Eq. (\ref{2.1}) is the tunneling Hamiltonian
\begin{equation}\label{2.4}
H_{\rm tun}= \sum_{\beta\kappa}\sum_{i}\left[
W^{}_{\beta,i\kappa}(t) a^\dag_{\beta\kappa}d^{}_{i} + {\rm
H.c.}\right] .
\end{equation}

\begin{figure}[ht]\begin{center}
  \includegraphics[width=6cm,angle=0]{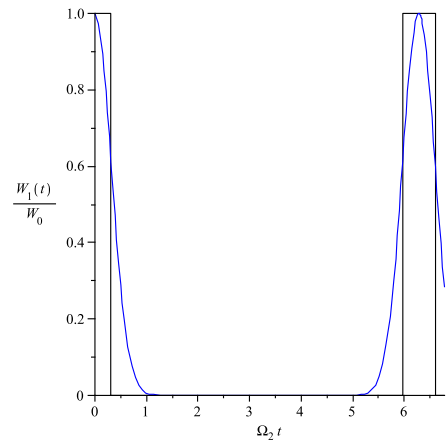}
  \caption{(Color online) Periodic tunneling potential in sudden and Gaussian
approximation}\label{f.1}
\end{center}
\end{figure}

Due to periodic shuttling both wells of the TS enter tunneling
contact with both banks. In the turnstile regime the time-dependent
tunneling potential has the form of periodic pulses, $ W(t) = \sum_p
w(t+ pT_2). $ In a rough approximation these pulses have a
rectangular shape of duration $T_1$ coming with the period $T_2$,
\begin{equation}\label{2.16s}
w(t) = w_0\left[\theta(t+T_1/2)-\theta(t-T_1/2)\right].
\end{equation}
This idealized picture of sudden approximation may be improved: if
the real time-dependent tunnel integral for rotating TS is
calculated, then the rectangular pulses transform into Gaussians,
\begin{equation}\label{2.16}
w(t) = w_0\exp \left[-\frac{t^2}{2(T_1/2)^2} \right].
\end{equation}
(see Fig. \ref{f.1}).

\begin{figure}[ht]\begin{center}
  \includegraphics[width=6cm,angle=0]{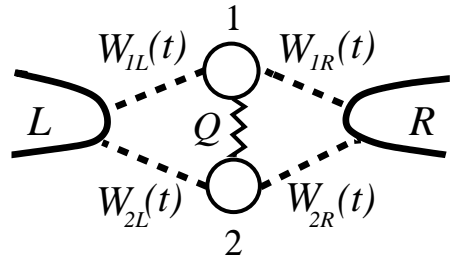}
  \caption{Double dot with time-dependent tunneling rate.
  Electrostatic interaction $Q$ prevents simultaneous occupation
  of both dots. }\label{f.3}
\end{center}
\end{figure}

Electron cotunneling through moving turnstile may be mimicked by
means of time dependent tunneling through an immovable two-channel
double dot shown in Fig. \ref{f.3}.
The appropriate time-dependence of the parameters $W_{li}(t)$ in
this construction is provided by the time-dependent gate voltages
regulating the magnitude of these parameters in such a way that in
the first half-period the channels $1L$ and $2R$ are open and the
channels $1R$ and $2L$ are blocked. As a result we get periodic in
time tunnel integrals, containing two sequences of pulses shifted by
half a period (see below).

\section{Single electron shuttling in periodic regime}

We formulate the periodic time-dependent problem in the following
way (see Fig. 2). Let the infinitesimally small bias voltage be
directed from the left bank to the right bank. At some moment, say
$t=0$, the {\em singly occupied} shuttle with an electron captured
in the site 2 is in the equilibrium position with the rotation angle
$\varphi=0$. Then within a quarter period $t=T_2/4$ shuttle rotates
to the angle $\varphi=-\pi/2$, the electron leaves well 2 for the
right bank, thus releasing the Coulomb blockade $U_{12}$, so that
the next electron from the left bank is allowed to enter the well 1.
At $t=3T_2/4$ the rotation angle is $\varphi=\pi/2$, and cotunneling
process is allowed again. At $t=T_2$ the shuttle loaded with one
electron returns into initial position with $\varphi=0$ and an
electron in the site 2 and the next cycle begins. Two electrons are
shuttled from the left lead to the right one during one cycle.

This picture implies that only two charge states of TS are
involved in cotunneling cycle, namely the one-electron states
$|\Lambda\rangle=|10\rangle$ and $|\Lambda\rangle=|01\rangle$
where an electron occupies the site 1 or 2, respectively. Two
other states, namely the state with empty wells
$|\Lambda\rangle=|00\rangle$, and the doubly occupied state
$|11\rangle$ appear only as virtual intermediate states. For the
sake of simplicity we assume that the state $|11\rangle$ is
completely suppressed by the Coulomb blockade potential $U_{12}$.
It is worthwhile to rewrite the Hamiltonian (\ref{2.1}) in terms
of the operators $X^{\Lambda\Lambda'}=|\Lambda\rangle\langle
\Lambda'|$. Then the first term acquires the form
\begin{equation}\label{2.2a}
H_{\rm sh} =\sum_{\Lambda=0,1,2}
E_{\Lambda}X^{\Lambda\Lambda}
\end{equation}
Here notation $|0\rangle =|00\rangle,\, |1\rangle = |10\rangle,\,
|2\rangle = |01\rangle$ is used.  We consider a symmetric shuttle,
where $E_1 = E_2$ is the energy of a singly occupied shuttle and
$E_0 = 0 $ is the energy of the empty TS. The tunneling Hamiltonian
reads
\begin{equation}\label{2.3a}
H_{\rm tun} = \sum_{\beta\kappa}\sum_{i=1,2}\left[ W^{}_{\beta i}(t)
a^\dag_{\beta\kappa}X^{0i} + {\rm H.c.}\right] .
\end{equation}
Four tunneling parameters, which are assumed to be
$\kappa$-independent are grouped in two periodic series shifted by
half period,
\begin{eqnarray}\label{2.5}
&&W_{L1}(t)=W_{R2}(t)=W_a(t) \nonumber \\
&&W_{L2}(t)=W_{R1}(t)=W_b(t)=  W_a(t- T_2/2),
\end{eqnarray}
so that
\begin{eqnarray}\label{2.17}
W_a(t)& = &\sum_{p=-\infty}^{+\infty} w(t+pT_2), \nonumber\\
W_b(t)&= & \sum_{p=-\infty}^{+\infty} w[t+(2p- 1)T_2/2)]
\end{eqnarray}
with $w(t)$ defined in Eq. (\ref{2.16s}) or (\ref{2.16}) ($p$ is
integer).

In order to describe the shuttling mechanism shown in Fig. \ref{f.2}
in mathematical terms, we  represent the Hamiltonian $H_{\rm tun}$
in the form of discrete Fourier series. For this sake we expand the
tunneling parameters (\ref{2.5}):
\begin{eqnarray}\label{2.6}
    W_a(t)&=& \sum\limits_{n=0}^\infty W_n \cos(\Omega_nt) \\\nonumber
    W_b(t)&=&
\sum\limits_{n=0}^\infty (-1)^n W_n \cos(\Omega_nt)
\end{eqnarray}
where $(\Omega_n=n\Omega=2\pi n/T_2)$. The Fourier coefficients for
two types of tunneling pulses (\ref{2.16s}), (\ref{2.16}) are
\begin{equation}\label{transr}
W_0 = w_0\tau,~~ W_n = \frac{2w_0\sin\pi n \tau}{\pi n}
\end{equation}
for the rectangular pulses and
\begin{equation}\label{transg}
W_0 = \sqrt{\frac{\pi}{2}}w_0\tau,~~ W_n = \sqrt{2\pi}w_0\tau
\exp\left[-\frac{(\pi n \tau)^2}{2}\right]
\end{equation}
for the Gaussian pulses. Here $\tau=T_1/T_2.$

Then inserting (\ref{2.6}) in (\ref{2.3a}) we see that the tunneling
Hamiltonian
\begin{eqnarray}\label{2.7}
H_{\rm tun} =  H_{\rm tun}^{(e)} +  H_{\rm tun}^{(o)},
\end{eqnarray}
is separated into two parts
\begin{eqnarray}\label{twoparts}
& H_{\rm tun}^{(e)} = \sum_\kappa W^{(e)}(t) a^\dag_{e\kappa} X_{e}
+{\rm H.c.}, \nonumber
\\
  \\
& H_{\rm tun}^{(o)} = \sum_\kappa W^{(o)}(t) a^\dag_{o\kappa} X_{o}+
{\rm H.c.} \nonumber
\end{eqnarray}
containing the contributions of the even and odd harmonics,
\begin{eqnarray}\label{evenodd}
W^{(e)}(t)= \sum_{m=0}^\infty 2 W_{2m}\cos(\Omega_{2m} t),\nonumber
\\ \\
W^{(o)}(t) = \sum_{m=0}^\infty 2 W_{2m+1}\cos(\Omega_{2m+1} t).
\nonumber
\end{eqnarray}

Even and odd creation operators for the lead and dot electrons are
defined as
\begin{eqnarray}\label{psi}
&&a^{\dag}_{e\kappa} = \frac{1}{\sqrt{2}}(a^{\dag}_{\kappa L} +
a^{\dag}_{\kappa R}),~ a^{\dag}_{o\kappa} =
\frac{1}{\sqrt{2}}(a^{\dag}_{\kappa L} -
a^{\dag}_{\kappa R}), \nonumber \\
\\
&& X^\dag_e = \frac{1}{\sqrt{2}}(X^{10}+X^{20}),~
   X^{\dag}_o = \frac{1}{\sqrt{2}}(X^{10}-X^{20})\nonumber
\end{eqnarray}
which implies
\begin{eqnarray}\label{2.10}
X^\dag_e X^{}_e= X^{ee}=\frac{1}{2}(X^{11} + X^{22} + X^{12} +
X^{21}),\nonumber
\\
X^\dag_o X^{}_o= X^{oo} = \frac{1}{2}(X^{11} + X^{22} - X^{12} -
X^{21}),\nonumber
\\
X^\dag_eX^{}_o=
X^{eo}=\frac{1}{2}(X^{11}-X^{22}-X^{12}+X^{21})\nonumber,
\\
X^\dag_oX^{}_e= X^{oe} = \frac{1}{2}(X^{11} - X^{22} + X^{12} -
X^{21})
\end{eqnarray}
The Hamiltonian of symmetric shuttle may be also rewritten in these
variables as
\begin{equation}\label{2.7d}
H_{\rm sh} = E_d( X^{ee}+ X^{oo})
\end{equation}
with $E_d=E_1=E_2$. A similar transformation for the symmetric leads
$(\varepsilon_{\kappa L}=\varepsilon_{\kappa R}=\varepsilon_{\kappa
})$ transforms the Hamiltonian (\ref{2.3}) into
\begin{equation}\label{2.7b}
H_{\rm lead} = \sum_{\kappa}\varepsilon_\kappa(a^\dag_{e\kappa}
a^{}_{e\kappa} +a^\dag_{o\kappa}a^{}_{o\kappa})~.
\end{equation}

As a result the Fourier transformed Hamiltonian of periodic
shuttling is reduced to  the system (\ref{2.7}), (\ref{2.7d}),
(\ref{2.7b}). Any Fourier harmonic $W_n$ is a result of averaging
the sequence of periodic pulses $w(t)$ with the corresponding
exponent. In the case of sudden rectangular pulses the convergence
of Fourier harmonics (\ref{transr}) is very poor, so that the whole
series should be taken into account, and even in this case the
problem of convergence known as the Gibbs defect remains. In the
case of more realistic Gaussian pulses the Fourier harmonics
(\ref{transg}) fall exponentially, so the convergence of the series
in the Hamiltonian (\ref{2.7}) is much better,
$$
W_{n+2}/W_n = \exp[-\pi^2\tau^2(2n+2)]
$$
The parameter $\tau$ may be estimated in a simplified model, where
the torque mode is imitated by rotation of a sphere with the radius
$d$ along the orbit with the radius $R$, so that the distance
between the leads equals $2(R+d)$. In this approximation $\tau=
d/\pi\sqrt{R(d+R)}$, so that the dwelling time $T_1$ of the shuttle
near the bank is controlled only by the geometrical factor $d/R$. In
the torque regime the interval $T_1$ is longer because the turnstile
slows down around the turning point

\subsection{Time-dependent canonical transformation for shuttling Hamiltonian}

Having in mind the approximate adiabaticity of shuttling
transport, we use the method of time-dependent canonical
transformation \cite{CFG2} for derivation of contunneling
Hamiltonian as well as for calculation of current operator (see
also Ref. \onlinecite{KF79}, where the time-independent canonical
transformation diagonalizing the Anderson impurity Hamiltonian in
the mean-field approximation was proposed).

As is shown in Ref. \onlinecite{CFG2}, the tunneling terms may be
ousted from the time-dependent Schr\"odinger operator ${\cal L} =
-i\hbar \partial/\partial t + H$ with $H = H_e+H_o$ defined in Eqs.
(\ref{2.7}) -- (\ref{2.7d}) by means of the transformation matrix
$e^{S_e(t) + S_o(t)}$ applied to the field operator $A^\dag$ with
components (\ref{psi}). To eliminate the tunneling terms from the
Shr\"odinger equation defined by the operator
\begin{equation}\label{2.13}
\widetilde{\cal L} =  -i\hbar \frac{\partial}{\partial t} +
\widetilde H~,
\end{equation}
one should take $\widetilde H$ in the following form
\begin{eqnarray}\label{canon}
\widetilde H = e^{S_e + S_o}(H_e+ H_o)e^{-S_o - S_e} + S_1 ,\nonumber \\
S_1 = i\hbar\int_0^1 d\lambda e^{\lambda (S_e + S_o)}(\dot{S}_e +
\dot{S}_o)e^{-\lambda (S_o + S_e)}
\end{eqnarray}
with
\begin{eqnarray}\label{2.12}
&& S_p= \sum_\kappa \left(u_{p\kappa} X^\dag_p a^{}_{p\kappa} -
{\rm H.c.} \right)
\end{eqnarray}
($p=e,o$). The coefficients $u_{p \kappa}$ are looked for in the
form
\begin{eqnarray}\label{2.14}
&& u_{e\kappa}(t) = \widetilde{u}_{e\kappa}(t) + i
\widetilde{v}_{e\kappa}(t) \nonumber
\\
\\
&& u_{o\kappa}(t) = {\widetilde{u}}_{o\kappa}(t) + i
{\widetilde{v}}_{o\kappa}(t). \nonumber
\end{eqnarray}
where the functions $\widetilde{u}_{e\kappa}(t)$ and
$\widetilde{u}_{o\kappa}(t)$ contain respectively even and odd
cosine harmonics, similarly to Eq. (\ref{evenodd}). The nonadiabatic
contributions $\widetilde{v}_{e\kappa}(t)$ and
$\widetilde{v}_{o\kappa}(t)$ contain correspondingly even and odd
sine harmonics.

The procedure of canonical transformation is described in detail in
Ref. \onlinecite{CFG2}. It is based on the Baker-Hausdorff expansion
of exponential operator. The coefficients ${u}_{p\kappa}^{(n)}$ and
${v}_{p\kappa}^{(n)}$  are found from the condition of compensation
of the term $H_{\rm}$ in the expanded Hamiltonian $\widetilde H$.
Since the rotating turnstile is singly occupied at any moment, we
project the effective Hamiltonian onto the subspace $\langle
e,o|\ldots|e,o\rangle$.

After elimination of $H_{\rm tun}$ both the shuttle Hamiltonian
$\widetilde H_{\rm sh}$ and the lead Hamiltonian $\widetilde H_{\rm
lead}$ acquire corrections both time-independent and oscillating.
Constant correction to $\widetilde H_{\rm sh}$ having the form of
level renormalization can be calculated exactly,\cite{CFG2} whereas
the search of nonzero Fourier harmonics defined by Eqs.
(\ref{2.13})-(\ref{2.14}) is rather complicated procedure.

It can be perceived that the time derivative in the operator ${\cal
L}$ (\ref{2.13}) generates terms containing the factor
$\Omega/\Delta_\kappa$, where $\Delta_\kappa = \varepsilon_\kappa -
E_d $. In what follows the terms $\widetilde{v}_{e\kappa}(t)$ and
$\widetilde{v}_{o\kappa}(t)$ proportional to this small factor will
be neglected.

Then in the leading order
\begin{equation}\label{canon-coeff}
u_{p\kappa} = - \frac{1}{\Delta_\kappa} W^{(p)}(t)
\end{equation}
with $p = e,o$. The time-dependent corrections to $H_{\rm lead}$
acquire the form of effective cotunneling Hamiltonian
\begin{eqnarray}
&&\widetilde H_{\rm lead} = H_{\rm lead} + H_{\rm cotun}(t)
\end{eqnarray}
These correction terms may be interpreted as a time-dependent
generalization\cite{GoldAv,KNG} of the well known Schrieffer-Wolff
transformation.\cite{SW}
\begin{eqnarray}\label{2.15}
H_{\rm cotun}(t)&=& \sum_{\kappa\kappa'}[ J^{ee}_{\kappa\kappa'}
a^\dag_{e\kappa}a^{}_{e\kappa'}  X^{ee}+ J^{oo}_{\kappa\kappa'}
a^\dag_{o\kappa}a^{}_{o\kappa'}  X^{oo} \nonumber \\
& + & J^{eo}_{\kappa\kappa'} a^\dag_{e\kappa}a^{}_{o\kappa'}
X^{oe} +{\rm H.c.} ]
\end{eqnarray}
where
$$
J^{pp'}_{\kappa\kappa'} = W^{(p)}(t) W^{(p')}(t)
\left(\frac{1}{\Delta_\kappa} + \frac{1}{\Delta_{\kappa'}}\right)
$$
are the components of time-dependent reflection and transmission
amplitudes (see below). Returning back to the original variables for
the leads $L,R$,  we transform (\ref{2.15}) into $H_{\rm cotun}(t) =
H_{\rho} + H_{\tau}$ with
\begin{eqnarray}\label{refl}
&& H_\rho(t) = H_{LL}+ H_{RR} = \\
&&\frac{1}{2}\sum_{\kappa\kappa'}( J^{ee}_{\kappa\kappa'}X^{ee}
+J^{oo}_{\kappa\kappa'}X^{oo})
(a^\dag_{L\kappa}a^{}_{L\kappa'}+a^\dag_{R\kappa}a^{}_{R\kappa'})\nonumber - \\
&&\frac{1}{2}\sum_{\kappa\kappa'}(J^{eo}_{\kappa\kappa'}X^{oe}
+J^{oe}_{\kappa\kappa'}X^{eo})(a^\dag_{L\kappa}a^{}_{L\kappa'}-
a^\dag_{R\kappa}a^{}_{R\kappa'})\nonumber
\end{eqnarray}
\begin{eqnarray}\label{tran}
&& H_\tau(t) = H_{LR}+ H_{RL} = \\
&&\frac{1}{2}\sum_{\kappa\kappa'}( J^{ee}_{\kappa\kappa'}X^{ee}
-J^{oo}_{\kappa\kappa'}X^{oo}) (a^\dag_{L\kappa}a^{}_{R\kappa'} +
a^\dag_{R\kappa}a^{}_{L\kappa'})+ \nonumber
\\
&&\frac{1}{2}\sum_{\kappa\kappa'} (J^{eo}_{\kappa\kappa'}X^{oe}
-J^{oe}_{\kappa\kappa'}X^{eo})(a^\dag_{L\kappa}a^{}_{R\kappa'}-
a^\dag_{R\kappa}a^{}_{L\kappa'})\nonumber
\end{eqnarray}
When deriving this Hamiltonian, multiplication rules (\ref{2.10})
have been used.

Before turning to calculation of the tunneling current, we simplify
the effective Hamiltonian in order to retain only the terms
responsible for the electron cotunneling though the turnstile.
First, we omit the term $H_\rho(t)$ (\ref{refl}) containing creation
of electron-hole pairs in the left and right leads. These terms are
essential in the problems where the infrared orthogonality
catastrophe modifies the current (Kondo or edge singularity
mechanism). In the absence of such channels they yield only higher
order corrections to the shuttling current. The remaining term
$H_\tau(t)$ rewritten in the original variables reads
\begin{widetext}
\begin{eqnarray}\label{hamilton}
H_\tau(t) &=& \frac{1}{4}\sum_{\kappa\kappa'}
\left[(J^{ee}-J^{oo})(X^{11}+X^{22})+(J^{ee}+J^{oo})(X^{12}+X^{21})\right]
(a^\dag_{L\kappa}a^{}_{R\kappa'} + a^\dag_{R\kappa}a^{}_{L\kappa'})
\nonumber
\\
&+& \frac{1}{4}\sum_{\kappa\kappa'} (J^{eo} + J^{oe})(X^{12} -
X^{21}) (a^\dag_{L\kappa} a^{}_{R\kappa'} - a^\dag_{R\kappa}
a^{}_{L\kappa'})
\end{eqnarray}
\end{widetext}
Since only the states near the Fermi level with $\varepsilon_\kappa
\sim \varepsilon_F$ contribute to the current at small enough
$\Omega$, we neglect the $\kappa\kappa'$ dependence in the exchange
integrals.

By construction the difference $J^{ee} - J^{oo}$ stems from
interlacing of the components $W_a(t)$ and $W_b(t)$ in the SW
transformation. Namely, the $2m$-th harmonic of this difference is
\begin{equation}\label{2.21}
J^{ee} - J^{oo} =
\frac{4}{\Delta}\sum_{m=0}^{\infty}\overline{W_a(t)W_b(t)\cos
\Omega_{2m}t}\cos \Omega_{2m}t,
\end{equation}
where the bar denotes averaging over the shuttling period. One can
check by inspection of Eqs. (\ref{2.17}), (\ref{2.16}) and Fig.
\ref{f.1} that these averages are exponentially small.
Correspondingly, the sum $J^{ee} + J^{oo}$ contains only the
averaged products $\overline{W^2_a(t)} = \overline{W^2_b(t)}$. These
averages give the main contribution to the electron shuttling and we
keep them in the tunneling current. Below we will be interested only
in a few first harmonics of tunneling current, which stem from the
cotunneling parameters
\begin{eqnarray}\label{3.1}
 &&J^{ee} = \frac{8 W_0^2+ 4W_2^2}{\Delta}+\frac{
16 W_0W_2}{\Delta}\cos 2\Omega t, \nonumber\\
&&J^{oo}=\frac{4W_1^2}{\Delta}(1+ \cos 2\Omega t), \\
&&J^{eo} = \frac{8W_0W_1 + 4W_1W_2}{\Delta}\cos \Omega t. \nonumber
\end{eqnarray}
In these expressions only zero to second harmonics in Eq.
(\ref{evenodd}) are taken into account. The higher orders terms
$\sim \cos n\Omega t$ may be obtained within the same procedure.

The three lowest harmonics of the tunneling potential $W(t)$
contribute also to the diagonal part of the transformed shuttle
Hamiltonian
\begin{equation}
\widetilde H_{\rm sh} = \widetilde{E}_d^e X^{ee}+\widetilde{E}_d^o
X^{oo}
\end{equation}
with
\begin{eqnarray}\label{2.18}
&&\widetilde E_d^p(t) = E_d - D_p(t)
\sum_\kappa\frac{1}{\Delta_\kappa},\nonumber\\
&&D_e =4 W_0^2+ 2W_2^2 + 8 W_0W_2\cos2\Omega
t\nonumber\\
&&D_o = 2W_1^2(1+\cos 2\Omega t).
\end{eqnarray}

The periodic in time perturbation (\ref{2.18}) results in a
reconstruction of the energy spectrum of the shuttle, which may be
described in terms of quasienergies, $f_m = E_d^p + m\Omega$ (see,
e.g., Ref. \onlinecite{Z73}). These states are also involved in the
tunneling current, but the corresponding tunneling channels open
only in the higher than fourth order correction in $W_{\beta i}$.

\section{Electron transport through a turnstile shuttle}

We start discussion of shuttling mechanism with general remarks.
It should be emphasized that the TS can transport electrons
preferably in one direction only provided its mirror symmetries $1
\leftrightarrow 2$ and/or $L\leftrightarrow R$ are slightly
violated. Indeed, if the state $|2\rangle$ with the electron in
the site 2 is chosen as the initial state, and the cycle shown in
Fig. \ref{f.2} is realized, then the electron charge is
transported from the left to the right. If the configuration
$|1\rangle$ is chosen as the initial state, then the same cycle
results in the electron transport from the right to the left.
Another way to change the current direction is to shift to $\pi$
the phase of torque vibration. Certainly, two directions of
single-electron cotunneling are equivalent if the mirror symmetry
is perfect, so that the charge transport through the TS arises as
a response to violation of this symmetry, whatever is the
microscopic mechanism of this violation.

To find the tunnel current through the turnstile, one should
calculate the time-dependent probabilities of the transitions from
the initial state at $t=t_0$ to the states with $1,2,\ldots 2M$
electron-hole pairs in the leads at $t= MT_2$ in accordance with
Fig. \ref{f.1}. In the problems of this type, one should distinguish
between the weak coupling and strong coupling regimes. In the weak
coupling regime the tunnel conductance is a superposition of elastic
ZBA and the set of inelastic replica, in which the finite bias
anomalies (FBA) arise because of excitations of higher harmonics of
the periodic coupling strength (or higher Floquet states in terms of
quasienergies).\cite{Shirl65,Z73} In the strong coupling regime the
whole Floquet spectrum is involved in the time evolution of the
system, determined by the evolution operator
\begin{equation}\label{4.1}
 U(t_0,t)=|\Psi(t)\rangle\langle\Psi(t_0)| .
\end{equation}
This problem is quite complicated because the highly excited Floquet
states belonging to continuum are involved in tunneling and the
relaxation channels, which open due to Auger-type processes of
multiple creation of electron-hole pairs with finite energy and
accompany the electron cotunneling. Leaving this regime for future
studies, we consider here only the perturbative regime.

The current operator is defined as
\begin{equation}
\hat I = e\frac{1}{2}(\dot{N}_R - \dot{N_L})
\end{equation}
where $\dot{N}_\beta =\frac{d}{dt}(\sum_{\kappa}
a^\dag_{\beta\kappa}a^{}_{\beta\kappa})$. Calculation of these
derivatives by means of equations of motion gives for $\hat I$
following equation
\begin{widetext}
\begin{eqnarray}\label{4.11a}
\hat I &=& \frac{ie}{4\hbar}\sum_{\kappa,\kappa'} \left(J^{ee}+
J^{oo}\right)\left(X^{12}+
X^{21}\right)\left(a^\dag_{L\kappa'}a^{}_{R\kappa}-
a^\dag_{R\kappa'}a^{}_{L\kappa}\right)\nonumber\\
\\
& +
&\frac{ie}{4\hbar}\sum_{\kappa,\kappa'}\left(J^{eo}+J^{oe}\right)
\left(X^{12}-X^{21}\right)\left(a^\dag_{L\kappa'}a^{}_{R\kappa}+
a^\dag_{R\kappa'}a^{}_{L\kappa} \right).\nonumber
\end{eqnarray}
\end{widetext}
Having in mind the exponential smallness of the terms $\propto
(J^{ee}-J^{oo})$ discussed in the previous section  we have kept in
this equations only the terms corresponding to the processes shown
in Fig. \ref{f.2}, where the electron transfer from the left to the
right bank or vice versa is accompanied by repopulation of the two
"seats" in the shuttle.

We calculate the tunneling conductance $G(\omega)$ arising as a
response to weak periodic in time bias $eV(t)$ by means of the
Kubo-Greenwood formula
\begin{equation}\label{4.13}
G(\omega,t)= \frac{-i}{\hbar\omega}\left[D(t,\omega) -
D(t,\omega=0)\right]
\end{equation}
Here retarded Green function $D(t,t')$ and its Fourier transform are
defined as
\begin{eqnarray}\label{3.2}
D(t,t')&=&-i\langle G| [\hat I(t), \hat I(t')]|G\rangle\Theta(t-t')
\nonumber\\
D(t,\omega)&=&\int d(t'-t) D(t,t')e^{-i\omega (t'-t)},
\end{eqnarray}
The initial "pure" state $|G\rangle$ is defined as a state with one
electron in one of the two sites (say, 2) and the filled Fermi
spheres in the bank electron reservoirs. This state is doubly
degenerate:
$$
|G\rangle_{e,o} = |\bar b\rangle|\left(|1\rangle
\pm|2\rangle\right)/\sqrt{2}
$$
Here $|\bar b\rangle$ stands for the filled Fermi spheres in the
leads. When choosing the Kubo-Greenwood equation in the form
(\ref{4.13}), we have taken into account the fact that the
time-dependent terms are present both in the Hamiltonian of
unperturbed itinerant electrons in the two banks (\ref{2.3}) with
applied bias
\begin{equation}\label{4.3}
H_b(t) = H_R + H_L -\mu(N_R + N_L)- eV(t)\frac{N_R - N_L}{2}.
\end{equation}
and in the tunneling Hamiltonian $H_{tun}$.

The leading time independent contribution to the current comes from
the $n=0$ harmonic of the shuttling current (\ref{4.11a}) and is
proportional given by the time-independent component $\sim
J^{ee(0)}/4$ in the first term of Eq. (\ref{4.11a}). By means of Eq.
(\ref{4.13}) and (\ref{hamilton}) we obtain the conventional
expression for the zero bias conductance
\begin{equation}\label{4.2}
G_0 = \frac{2\pi e^2}{\hbar}\rho^2_F|J^{ee(0)}/4|^2,
\end{equation}
where $\rho_F$ is the electron density of states in the leads. The
prefactor 2 in this spinless model appeared due to the specific
charge transfer mechanism in a turnstile shuttle: unlike the
standard shuttle with a single "seat", the two-seat shuttle
transfers two electrons within a period, as is seen from Fig.
\ref{f.2}.

Equation (\ref{4.2}) determines the time-averaged background of a
tunnel current through a turnstile. It is important that the
coupling constant $J^{ee(0)}$ contains the maximal values of
tunneling matrix elements $w_0$ corresponding to the closest contact
between the shuttle and metallic banks, although the coupling
strength is essentially weakened due to the factor $\tau$ which
characterizes the dwelling time of the shuttle in the nearest
vicinity of the banks [see Eq. (\ref{transg})]. Taking into account
corrections given by the first harmonic $J^{oo(1)}$ we obtain the
average conductance
\begin{equation}
\overline{G} =\frac{\pi e^2}{8\hbar} \rho^2_F|J^{ee(0)}|^2[1+
4e^{-(\pi\tau)^2}] .
\end{equation}

To find full time-dependent conductance let us rewrite Eq.
(\ref{4.11a}) for the current operator in the form
\begin{widetext}
\begin{eqnarray}\label{4.5}
\hat I(t) &=& \frac{ie}{\hbar} \left [I_0(X^{12}+X^{21}) +
I_2(X^{12} + X^{21})\cos 2\Omega t\right]
\sum_{\kappa}\sum_{\kappa'} \left(a^\dag_{L\kappa'}a^{}_{R\kappa} -
a^\dag_{R\kappa'}a^{}_{L\kappa} \right) \nonumber\\
&\phantom{a}& + \frac{ie}{\hbar} I_1(X^{12}-X^{21})\cos \Omega t
\sum_{\kappa}\sum_{\kappa'} \left(a^\dag_{L\kappa'}a^{}_{R\kappa}+
a^\dag_{R\kappa'}a^{}_{L\kappa}\right)
\end{eqnarray}
\end{widetext}
with
\begin{eqnarray}\label{4.6}
I_0&=&\frac{2 W_0^2 + W_1^2 + W_2^2}{\Delta},
   \nonumber \\
I_1&=&\frac{4 W_0 W_1 + 2 W_1W_2 }{\Delta},\nonumber\\
I_2&=&\frac{W_1^2 + 4 W_0W_2 }{\Delta}.
\end{eqnarray}
Substitution of (\ref{transg}) yields
\begin{eqnarray}\label{4.6b}
I_0&=& \frac{\pi (w_0\tau)^2}{\Delta}[1 +
2e^{-y}+2e^{-4y}],\nonumber\\
I_1&=& \frac{4\pi (w_0\tau)^2}{\Delta}e^{-y/2}[1+
e^{-2 y}],\\
I_2&=& \frac{2\pi(w_0\tau)^2}{\Delta} e^{-y} [ 1+ 2e^{-y}].\nonumber
\end{eqnarray}
with $y=(\pi\tau)^2$ (see (\ref{2.16})). The parameter $\exp(-y)$
controls the smallness of the contribution of higher harmonics.

Operators $X^{12}$ describing elastic processes with zero frequency
do not contribute to the cotunneling dynamics and give the
combinations $ X^{11}\pm X^{22}$ in the current correlation
functions in (\ref{3.2}). After averaging these combinations yield
the factors 1 and 0, respectively in the resulting equations for the
conductance. Inserting Eqs. (\ref{4.5}),(\ref{4.6}) into Eq.
(\ref{3.2}) and then into Eq. (\ref{4.13}) we calculate the
electron-hole loops for itinerant carriers in the standard way and
get finally
\begin{equation}\label{4.7}
G(0,t)= \frac{e^2}{h}\rho_F^2\left[\left(I_0^2 -
\frac{I_1^2}{2}\right) + 2\left(I_0I_2 - \frac{I_1^2}{4}\right) \cos
2\Omega t \right]
\end{equation}

\begin{figure}[ht]\begin{center}
  \includegraphics[width=8cm,angle=0]{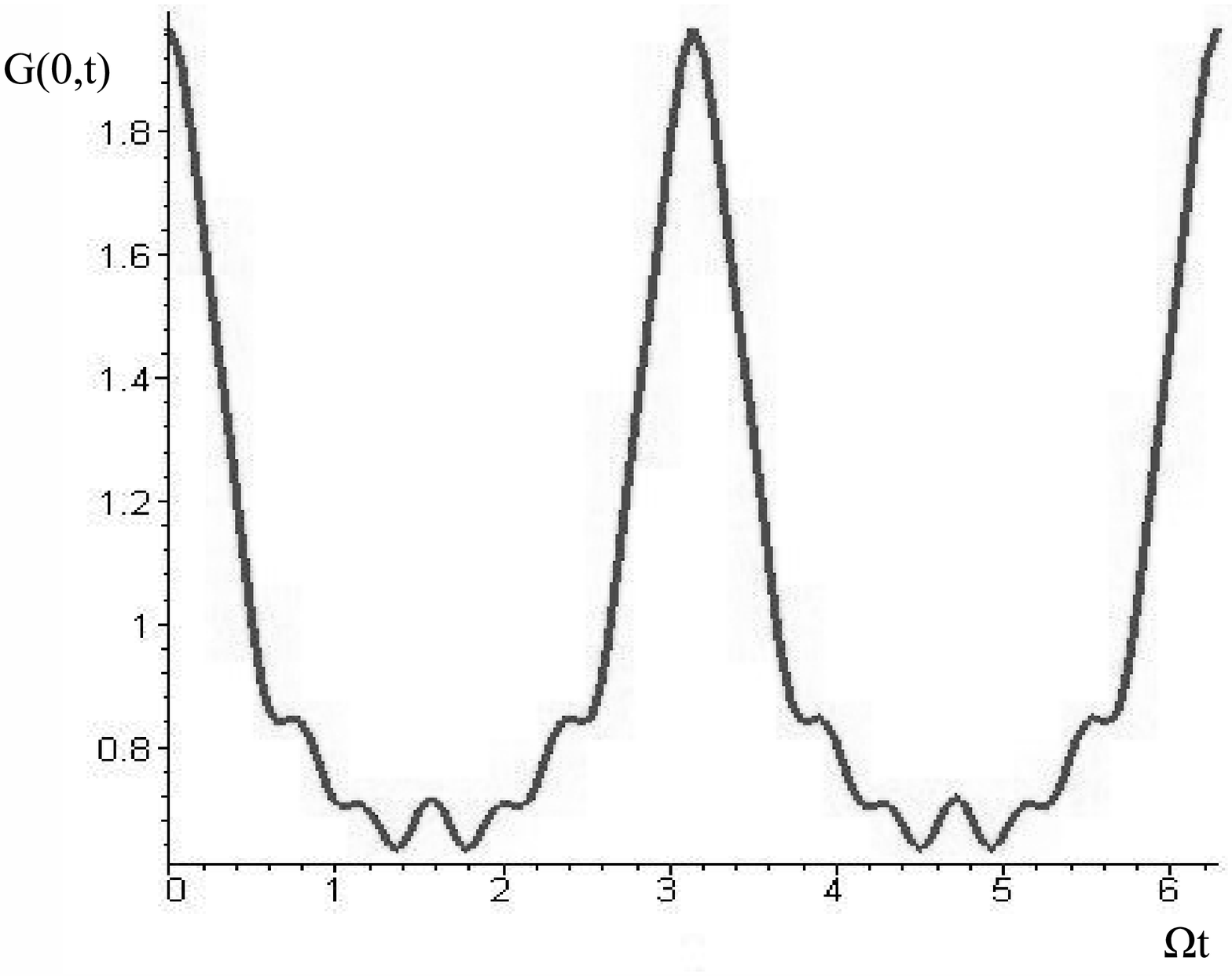}
  \caption{Time dependent zero-bias anomaly in oscillating conductance
$G(0,t)= \sum_n A_n\cos 2n\Omega t$ where for the sake of
demonstration we have chosen a constant ratio $A_{n+1}/A_n =1/2$.
Terms with $n\leq 4$ are taken into account.}\label{f.6}
\end{center}
\end{figure}

This equation describes adiabatic two-electron shuttling. We see
that the oscillating motion of a turnstile with the frequency
$\Omega$ results in an appearance of periodic component with the
frequency $2\Omega$ in the time-dependent conductance. Higher
harmonics $2m\Omega$ in the odd modes and $(2m+1)\Omega$ in the
coupling constants (\ref {3.1}) also may be taken into account.
These harmonics result in an appearance of the higher even harmonics
$\propto \cos 2n\Omega$ with smaller amplitudes in oscillating
conductance. An example of oscillating conductance including 4
harmonics is shown in Fig. \ref{f.6}.

In principle, we may consider nonadiabatic effects as well. One may
take into account weak non-adiabatic corrections, which stem from
the terms $\widetilde{v}_{e\kappa}(t)$ and
$\widetilde{v}_{o\kappa}(t)$ in Eq. (\ref{2.14}). These corrections
may be observed as phase shifts in the cosine functions.\cite{CFG2}
Truly non-adiabatic effects arise when the contact between the
shuttle and the banks results in reconstruction of the spectra of
electrons in the shuttle due to a fast enough periodic passage of
two islands forming a turnstile along the banks. Such reconstruction
may be treated in terms of
quasienergies,\cite{Shirl65,Z73,Baron77,Toka08} which are just a
realization of the Floquet theorem for the time-periodic systems. In
this case the energy conserves in the process of cotunneling under
the perturbation with the period $T_1= 2\pi/\Omega$ to within the
"Umklapp" processes, $E_f = E_i + m \Omega$, where $E_{i,f}$ are the
initial and final states of cotunneling act. As a result, the
satellites $\sim 2 n \hbar\Omega$ should arise in the conductance as
finite bias anomalies.  These satellites cannot be incorporated in
the above calculation scheme. Development of an appropriate
non-perturbative description is beyond the framework of the present
paper.

\section{Conclusions}

Our analysis of periodic shuttling within a time-dependent Anderson
model with periodic tunneling (\ref{twoparts}) perturbing odd and
even modes of the shuttle and bath subsystems has shown that the
shuttle works as a harmonic analyzer, which transforms the input
signal (\ref{evenodd}) into the output adiabatic signal (\ref{4.7}).
In the approximate equation (\ref{4.7}) only two first harmonics
$\Omega$ and $2\Omega$ are taken into account. Numerical estimates
presented in Fig. \ref{f.6} show that even in the case of large
enough parameter $\exp(-y)$ the higher harmonics only weakly perturb
the basic features of the effect, namely permanent background
$\propto I^2_0$, where the magnitude of cotunneling strength is
controlled by the parameter $\tau^4$ [see Eq. (\ref{4.6b}), and the
periodic temporal oscillations with the leading harmonic $2\Omega$.

We have found three realizations of this model, namely the turnstile
suspended on an elastic string in two configurations (Fig.
\ref{f.0}) and a double quantum dot with gate-controlled tunneling
parameters $W_{\beta i}$ (Fig. \ref{f.3}). In principle turnstile
configurations may be realized also in molecular motors. In this
paper we considered only the charge transport induced by electrical
bias, but the ratchet-like turnstiles could also be proposed.


\begin{thebibliography}{99}

\bibitem{Thou} D. J. Thouless, Phys. Rev. B {\bf27}, 6083 (1983)

\bibitem{Gorel} L. Y. Gorelik, A. Isacsson, M. V. Voinova, B.
Kasemo, R. I. Shekhter, and M. Jonson, Phys. Rev. Lett. {\bf 80},
4526 (1998).

\bibitem{Poth} H. Pothier, P. Lafarge, C. Urbina, D. Esteve, and M.
H. Devoret, Europhys. Lett., {\bf17}, 249 (1992)

\bibitem{Swit} M. Switkes, C. M. Markus, K. Campman, and A.
Gossard, Science {\bf283}, 1905 (1999)

\bibitem{Zor} S. Lotkov, S. A. Bogoslovsky, A. B. Zorin, and J.
Nimeyer, Appl. Phys. Lett., {\bf78}, 946 (2001); Y. Ono and Y.
Takahashi, Appl. Phys. Lett. {\bf82} (2003) 1221.

\bibitem{Watson} S. K. Watson, R. M. Potok, C. M. Marcus, and V.
Umansky, Phys. Rev. Lett. {\bf 91}, 258301 (2003).

\bibitem{Aono} T. Aono, Phys. Rev. Lett. {\bf93}, 116601 (2004).

\bibitem{Gorev} R. I. Shekhter, L. Y. Gorelik, M. Jonson, Y. M.
Galperin, and M. V. Vinokur, J. Comput. Theor. Nanosci. {\bf4}, 860
(2007)

\bibitem{Erbe} A. Erbe, C. Weiss, W. Zwerger, and R. H. Blick,
Phys. Rev. Lett. {\bf 87} 096106 (2001).

\bibitem{Kot2004} D. V. Scheible, C. Weiss, J. P. Kotthaus, and R.
H.Blick, Phys. Rev. Lett.{\bf 93}, 186801 (2004); D. V. Scheible and
R. H. Blick, Appl. Phys. Lett. {\bf 84}, 4632 (2004).

\bibitem{Leroy} B. J. LeRoy, S. G. Lemay, J. Kong, and C. Dekker,
Nature {\bf432}, 371 (2004).

\bibitem{Sheht09} R. I. Shekhter, F. Santandrea, G. Sonne, L. Y.
Gorelik, and M. Jonson, Low Temp. Phys. {\bf 35}, 662 (2009).

\bibitem{Koenig} D. R. Koenig, E. M. Weig, and J. P. Kotthaus, Nature
Nanotechnology {\bf 3}, 482 (2008).

\bibitem{Koendis} D. R. Koenig, Ph.D. Thesis, L\"udwig-Maksimilian
Universit\"at, M\"unchen, 2008.

\bibitem{GR} L.I. Glazman and M.E. Raikh, Pis'ma Zh. Eksp. Teor.
Fiz. {\bf67}, 1276 (1988) [Sov. Phys. -- JETP Lett. \textbf{47}, 452
(1988)]
%
\bibitem{Ng} T. K. Ng, P. A. Lee, Phys. Rev. Lett. \textbf{61}, 1768 (1988).
%
\bibitem{Kogla} L. Kowenhoven and L. Glazman, Physics World {\bf14},
33 (2001).
%
\bibitem{KKSV} M. N. Kiselev, K. Kikoin, R. I. Shekhter, and V. M. Vinokur,
Phys. Rev. B {\bf 74}, 233403 (2006)
%
\bibitem{CFG2} G. Cohen, V. Fleurov, and K. Kikoin, Phys. Rev. B
{\bf79}, 245307 (2009).
%
\bibitem{Shirl65} J. N. Shirley, Phys. Rev. B{\bf138}, 979 (1965).
%
\bibitem{Z73} Y. B. Zeldovich, Uspekhi Fizicheskikh Nauk {\bf 110},
139 (1973) [Sov. Phys. Usp., {\bf 16}, 427 (1973)]
%
\bibitem{Baron77} S. R. Barone, M. A. Narowich, and F. J. Narowich, Phys.
Rev. A {\bf15}, 1109 (1977)
%
\bibitem{Toka08} N. Tsuji, T. Oka, and H. Aoki, Phys. Rev. B {\bf78}, 235124
(2008).
%
\bibitem{KF79} K. A. Kikoin and V. N. Fleurov, Sov. Phys. - JETP {\bf50}, 535 (1979)
[Zh. Eksp. Teor. Fiz. {\bf77}, 1062 (1979)].

\bibitem{GoldAv} Y. Goldin and Y. Avishai, Phys. Rev. B {\bf61},
16750 (2000).

\bibitem{KNG} A. Kaminski, Yu. V. Nazarov, and L.I. Glazman, Phys.
Rev. B {\bf 62}, 8154 (2000)

\bibitem{SW} J. R. Schrieffer and P. A. Wolff, Phys. Rev. {\bf 149}, 461
(1966).


\end{thebibliography}
\end{document}